\documentclass[aps,superscriptaddress,nofootinbib,eqsecnum,prd,notitlepage,twocolumn,showkeys]{revtex4-1} 

\pdfoutput=1

\usepackage{amsfonts}
\usepackage{amsmath}
\usepackage{amssymb}
\usepackage{graphicx,color}
\usepackage{float}
\usepackage{hyperref}
\usepackage{subfigure}
\usepackage{dcolumn}
\usepackage{soul}
\usepackage{ulem}
\usepackage{verbatim}

\begin{document}

\title{Reassessing constant-roll Warm Inflation}
\author{Sandip Biswas}
\email{sandipb20@iitk.ac.in}
\affiliation{Department of Physics, Indian Institute of Technology, Kanpur: 208016,
Uttar Pradesh, India}

\author{Kaushik Bhattacharya}
\email{kaushikb@iitk.ac.in}
\affiliation{Department of Physics, Indian Institute of Technology, Kanpur: 208016,
Uttar Pradesh, India}

\author{Suratna Das}
\email{suratna.das@ashoka.edu.in}
\affiliation{Department of Physics, Ashoka University,
   Rajiv Gandhi Education City, Rai, Sonipat: 131029, Haryana, India}

\begin{abstract}

Departing from standard slow-roll conditions is one way of putting the inflationary paradigm to test, and constraining the dynamics of the inflaton field with a constant-rate of roll of the inflaton field, a.k.a. the constant-roll scenario, is one way of exploring such deviation from the standard slow-roll dynamics. In this manuscript we explore such a possibility in a variant inflationary scenario, known as Warm Inflation. We construct and derive the conditions for having constant-roll WI models where inflation lasts at least for 60 $e-$folds, gracefully exits the constant-roll inflation phase, and maintains near thermal equilibrium of the system which is an essential feature of WI in the slow-roll regime. We show that while certain models of WI (the ones with dissipative coefficient as a function of temperature alone) can accommodate constant-roll dynamics, others (with dissipative coefficient as a function of temperature and the inflaton field both) fail to maintain thermal equilibrium once the constant-roll condition is imposed and hence cannot produce a constant-roll WI phase.

\end{abstract}

\maketitle

\section{Introduction}

On the one hand, cosmic inflation \cite{Kazanas:1980tx, Guth:1980zm, Sato:1981ds, Sato:1980yn, Linde:1981mu, Albrecht:1982wi} solves the fine-tuning problems of the Big Bang cosmology, i.e. the horizon problem, the flatness problem and the monopole problem. On the other hand, it helps generate inhomogeneties in an otherwise homogeneous and isotropic Friedmann-Lemaitre-Robertson-Walker (FLRW) background, which eventually give rise to the large scale structures we see around us today as well as the fluctuations in the temperature of the Cosmic Microwave Background Radiation (CMBR) which have been measured to a very good accuracy in recent times. The dynamics of the inflationary phase is often portrayed as the dynamics of a single scalar field which is slowly rolling down its approximately flat potential. The slow-roll dynamics of the inflaton field is often invoked in any standard inflationary dynamics as it helps gracefully exit the inflationary phase as well as generate nearly scale-invariant spectra of primordial perturbations (both scalar and tensor), and thus makes definite predictions which can be tested through the observations of the CMBR and the large scale structures \cite{Riotto:2002yw, Baumann:2009ds, Mishra:2024axb}. This simple vanilla model of single field slow-roll inflation is in good agreement with the current data \cite{Planck:2018jri}. Significant departure from the slow-roll dynamics, as it happens during an ultra-slow-roll phase \cite{Kinney:2005vj, Dimopoulos:2017ged} for example, results in departure from the nearly scale-invariant spectra that slow-roll generates. It has been shown that a transient phase of Ultra-slow-roll yields large primordial perturbations which eventually collapse to form Primordial Black Holes \cite{Motohashi:2017kbs}, a coveted candidate for Dark Matter \cite{Carr:2016drx}. 

Despite being in accordance with the current observations, inflation is still a paradigm which allows one to explore many different ways to depart from the standard single field slow-roll dynamics of inflation while still being in tune with the available data. One such attempt is to consider constant-roll inflation \cite{Martin:2012pe, Motohashi:2014ppa} which, depending on the value of the constant-roll parameter $\beta$, can be significantly different from the standard slow-roll inflation. The concept of constant-roll inflation was first introduced in \cite{Martin:2012pe}, but the dynamics of constant-roll was first clearly penned down in \cite{Motohashi:2014ppa}. The term ``constant-roll'' was also first coined in \cite{Motohashi:2014ppa}. As has been indicated in \cite{Motohashi:2014ppa}, the name ``constant-roll'' signifies a constant rate of roll of the inflaton field, an additional constraint imposed on the dynamics of the inflaton field. The constraint can be written as follows \cite{Martin:2012pe}: 
\begin{eqnarray}
    \ddot\phi+3\beta H\dot\phi=0,
    \label{constant-roll-condition}
\end{eqnarray}
where the constant-roll parameter $\beta$ is a constant and can take both positive and negative values. Note that the constant-roll parameter $n$ in \cite{Martin:2012pe} is equal to $-3\beta$ in our case. The above equation can be rewritten as $d\ln\dot\phi/dN=-3\beta$, where $N$ is the number of $e-$folds. This clearly shows that  $\dot\phi$ changes with a constant rate ($-3\beta$) with $e-$folds, and hence the name. If $|\beta|\ll1$, then one can ignore $\ddot\phi$ with respect to the $3H\dot\phi$ term, and in such cases the dynamics boils down to the familiar slow-roll dynamics. On the other hand, $\beta=1$ corresponds to the ultra-slow-roll dynamics for a phase when $\partial V/\partial\phi=0$, where $V$ represents the inflaton potential \cite{Kinney:2005vj, Dimopoulos:2017ged}. One can get an estimate of the allowed values of $\beta$ in two different ways. First of all, it has been shown in \cite{Morse:2018kda,Lin:2019fcz} that only small-$\beta$ constant-roll models ($\beta<0.5$) can be universal attractors, whereas large-$\beta$ constant-roll ($\beta>0.5$) should be treated as an initial transient phase followed by an evolution dominated by slow-roll dynamics. The constant-roll parameter $\bar\eta$ in \cite{Morse:2018kda,Lin:2019fcz} is equal to $3\beta$ in our case. Secondly, the perturbations generated during a constant-roll inflationary phase (not the initial transient phase followed by slow-roll dynamics) are compatible with current data only when $\beta$ is negative and of the order $10^{-3}$ \cite{Motohashi:2017aob}. Note that the constant roll parameter $\beta$ in \cite{Motohashi:2017aob} is equal to $-3\beta$ in our case. However, it is shown in \cite{Gao:2019sbz} that $\beta\sim1$ is also allowed observationally, but such constant-roll inflation models fails to gracefully exit an inflationary phase. Note that the constant-roll parameter $\eta_H$ in \cite{Gao:2019sbz} is equal to $3\beta$ in our case. This indicates that despite imposing the constant-roll condition, given in Eq.~(\ref{constant-roll-condition}), on a single-field inflationary dynamics, to be an attractor solution as well as to remain viable by the data (in addition, to be able to gracefully exit an inflationary phase), the dynamics should not differ much from a slow-roll evolution.  

In this article we will reassess the constant-roll dynamics in a variant inflationary scenario, known as Warm Inflation \cite{Berera:1995ie}. Unlike in the standard single-field inflationary scenario described above, where the couplings of the inflaton field is generally ignored during an inflationary phase, in Warm Inflation (WI) these couplings play a significant role in dissipating inflaton's energy to maintain a constant radiation bath throughout inflation (for more recent reviews on WI, see \cite{Kamali:2023lzq, Berera:2023liv}). In doing so, WI can gracefully exit to a radiation dominated era, whereas the more standard inflationary scenario, which we will refer to as Cold Inflation (CI) henceforth, requires a subsequent reheating phase, dynamics of which is still largely unknown. Also, the requirement of reheating post inflation in CI scenarios restricts the classes of inflationary potentials to the ones with a minimum. Otherwise, one requires other mechanism to reheat the universe, such as gravitational reheating \cite{Ford:1986sy, Chun:2009yu} for example, apart from standard reheating dynamics via the inflaton's oscillation at the minima of the potential \cite{Mishra:2024axb}. As WI doesn't require a subsequent reheating phase, such scenarios can accommodate larger classes of inflaton potentials. For example, generalised runaway potentials, which do not have minima, have been explored in \cite{Das:2020xmh} and was shown that the WI dynamics smoothly transits to a radiation dominated era. However, graceful exit in WI is a much more complex process than in CI as has been pointed out in \cite{Das:2020lut}. Some of the attractive features of WI include (a) generation of smaller tensor-to-scalar ratio which helps accommodate potentials, like quartic and quadratic, which are otherwise ruled out in CI for generating way too much tensor perturbations compared to scalar perturbations \cite{Bartrum:2013fia}, \footnote{Recently two numerical codes have been developed to compute perturbations in WI \cite{Montefalcone:2023pvh, Ballesteros:2023dno}.} (b) generation of Primordial Black Holes without departing from the slow-roll dynamics \cite{Arya:2019wck, Bastero-Gil:2021fac, Correa:2022ngq, Arya:2023pod}, (c) accommodating the de Sitter Swampland Conjecture \cite{Ooguri:2018wrx, Garg:2018reu} naturally \cite{Das:2018hqy, Motaharfar:2018zyb, Das:2018rpg, Das:2019hto, Das:2019acf} which otherwise disfavours the standard CI dynamics \cite{Kinney:2018nny}, to name of few. 

An important feature of WI is that the radiation bath produced during WI is treated to be in near thermal equilibrium due to which a temperature $T$ can be assigned to the bath. The inflaton field, on the other hand, may or may not thermalize with the radiation bath. The radiation bath, produced during WI is sub-dominant (the potential energy density of the inflaton field dominates the energy density during WI just like it happens in CI), but not negligible. One measure to differentiate between CI and WI is that if the radiation energy density $\rho_r$ is greater (less) than the Hubble parameter $H$ during an inflationary phase, then it is to be treated as WI (CI) \cite{Kamali:2023lzq}. Upon thermalization of the radiation bath, the WI condition, $\rho_r^{1/4}>H$, translates into the thermalization condition $T>H$, which is to be maintained throughout WI. The near thermal equilibrium condition also plays a major role in determining the cosmological perturbations produced during WI. Unlike CI, where the primordial perturbations are quantum in nature, WI produces thermal fluctuations as well, which are classical and determines the form of the scalar power spectrum \cite{Kamali:2023lzq, Berera:2023liv}. It has been noted earlier that it becomes a daunting task to maintain near thermal equilibrium of the radiation bath during WI when the dynamics veers away from slow-roll dynamics, such as in the case of ultra-slow-roll \cite{Biswas:2023jcd}. It has been shown in \cite{Biswas:2023jcd} that the thermalization condition, which is an essential feature of WI, can be maintained only for a few $e-$foldings and that too in a very restrictive class of WI models where the dissipative coefficient is a function of both temperature and the inflaton field. 

Constant-roll dynamics in WI has been explored a few times in the literature before \cite{Kamali:2019wdh, Mun:2021kzb, Setare:2021iei, AlHallak:2021hwb, Saleem:2023aof}. The latter three works \cite{Setare:2021iei, AlHallak:2021hwb, Saleem:2023aof} deal with constant-roll WI in modified gravity theories, and thus of no interest for the present discussion. The first analysis \cite{Kamali:2019wdh} dealt with toy models of WI where the parameter $Q$ (ratio of the two frictional terms in the equation of motion of the inflaton field in WI) is kept constant. The second analysis \cite{Mun:2021kzb} dealt with some specific form of $Q$ depending on the inflaton field $\phi$ and the temperature $T$, though they did not consider the standard forms of dissipative coefficients appearing in various WI models \cite{Kamali:2023lzq, Berera:2023liv}. Their parametrization of $Q$ as a function of $\phi$ and $T$ are rather ad hoc. More importantly, both these papers do not take into account the thermalization condition which requires to be maintained even in a constant-roll dynamics of WI (simply to call it a WI scenario). Above all, both these works do not comment on graceful exit conditions in constant-roll WI. We will comment more on these works in a latter section. 

What we want in this paper is to obtain constant-roll WI models where (a) constant-roll defines the full dynamics of an inflationary phase (at least 60 $e-$folds of it), and not just a transient phase of it, (b) graceful exit from the constant-roll phase is ensured, and (c) near thermal equilibrium condition ($T>H$) is maintained throughout the constant roll phase so that we can logically call it a constant-roll WI model. In doing so, we will explore all possible kinds of WI scenarios, and will not restrict ourselves to only constant-$Q$ toy models of WI. We will explore both weak dissipative $(Q\ll1)$ and strong dissipative $(Q\gg1)$ scenarios of WI as well. We have furnished the rest of the paper as follows. In Sec.~(\ref{CR-CI}) we revisit the constant-roll CI scenario and determine the graceful exit conditions, which, to the best of our knowledge, has not been addressed so far in the literature. In doing so, we will show that the graceful exit itself constrain the constant-roll parameter $\beta$. In Sec~(\ref{CR-WI-graceful-exit}) we then determine the graceful exit conditions for any WI model. In Sec.~(\ref{CR-WI-thermal-equilibrum}) we will determine the conditions to maintain near thermal equilibrium during a constant-roll WI scenario. We will show that to maintain near thermal equilibrium condition in constant-roll WI scenario, one cannot differ much from a slow-roll evolution. This is a condition which was also obtained for constant-roll CI models to be compatible with the current data. However, it comes from the perturbation analysis of the constant-roll CI dynamics. Here, in the case of WI, even the background evolution constrains the dynamics to evolve closely to a slow-roll dynamics. We will also show that WI models where the dissipative coefficient is a function of temperature and the inflaton field, cannot maintain the thermalization condition during a constant-roll phase, and thus constant-roll dynamics cannot be realised in the class of such WI models. In Sec.(\ref{conclusion}), we will discuss the main results obtained in this paper and will conclude with some future outlooks.

\section{A fresh look at constant-roll Cold Inflation: Graceful exit condition}
\label{CR-CI}

In standard CI scenario, the background geometry evolves according to the first and second Friedmann equations, whereas the dynamics of the scalar inflaton field $\phi$ is governed by the Klein-Gordon equation. These three equations together determine the evolution of the universe during CI, which can be written as 
\begin{eqnarray}
&&3M_{\rm Pl}^2H^2=\frac{\dot\phi^2}{2}+V(\phi),\label{Hsquare}\\
&&2M_{\rm Pl}^2\dot H=-\dot\phi^2,\label{Hdot}\\
&&\ddot\phi+3H\dot\phi+V,_\phi=0\label{KG-eq},
\end{eqnarray}
respectively. Here the overdot denotes derivative with respect to the cosmic time $t$, $V,_\phi\equiv dV(\phi)/d\phi$, $H\equiv \dot a/a$ is the Hubble parameter determining the rate of change of the scale factor $a$ of the Friedmann-Lema\^{i}tre-Robertson-Walker background metric, and $M_{\rm Pl}$ is the reduced Planck mass. The evolution of the background is often quantified in terms of the Hubble slow-roll parameters defined as \cite{Martin:2012pe}
\begin{eqnarray}
\epsilon_{i+1}=\frac{d\ln\epsilon_i}{dN},
\end{eqnarray}
where $N\equiv\ln a$ is the number of $e-$folds. The first two Hubble slow-roll parameters are of interest to us, which starting from $\epsilon_0=1/H$ in the above relation, can be written, as
\begin{eqnarray}
\epsilon_1=-\frac{\dot H}{H^2},\quad\quad \epsilon_2=\frac{\dot\epsilon_1}{\epsilon_1 H}.\label{epsilons}
\end{eqnarray}
One requires $\epsilon_1<1$ to ensure an inflationary phase, and $\epsilon_1\sim1$ designates end of inflation. The standard slow-roll of the inflaton field requires the subsequent Hubble slow-roll parameters to be smaller than unity $(|\epsilon_i|\leq1)$. However, the subsequent slow-roll parameters can become of order unity or larger when the dynamics deviates from slow-roll. Even if the dynamics deviates from slow-roll, one requires the first Hubble slow-roll parameter to be smaller than unity $(\epsilon_1<1)$ to ensure inflation. Thus, to end an inflationary phase, $\epsilon_1$ should grow from smaller than unity values to unity. Therefore, graceful exit demands a growing $\epsilon_1$ during any inflationary phase, slow-roll or otherwise, which can be written as 
\begin{eqnarray}
\frac{d\epsilon_1}{dN}=\frac{\dot\epsilon_1}{H}>0.
\end{eqnarray}
As $H$ is always positive in an expanding Universe, we thus require $\dot\epsilon_1>0$ to gracefully exit any inflationary phase. 

Imposing the constant-roll constraint given in Eq.~(\ref{constant-roll-condition}), the Klein-Gordon equation in Eq.~(\ref{KG-eq}) would become 
\begin{eqnarray}
    3(1-\beta)H\dot\phi=-V,_\phi. \label{KG-CR}
\end{eqnarray}
We would first like to investigate for what values of $\beta$ constant-roll CI can make a graceful exit. From Eqs.~(\ref{epsilons}), we note that 
\begin{eqnarray}
    \dot\epsilon_1=-\frac{1}{H^2}\left(\ddot H-\frac{2\dot H^2}{H}\right). \label{dot-eps1}
\end{eqnarray}
Note that as this equation is derived from Hubble slow-roll parameter which depends only on the background evolution (and not on the form of the inflaton potential), this equation is valid for slow-roll or beyond slow-roll CI or WI. 

In constant-roll inflation, we get 
\begin{eqnarray}
M_{\rm Pl}^2\ddot H=3\beta H\dot\phi^2,
\end{eqnarray}
whereas, $\dot H$ can be read from Eq.~(\ref{Hdot}). Therefore, we obtain 
\begin{eqnarray}
    \ddot H-\frac{2\dot H^2}{H}=\frac{3H\dot\phi^2}{M_{\rm Pl}^2}\left(\beta-\frac{\dot\phi^2}{6M_{\rm Pl}^2H^2}\right).
\end{eqnarray}
It is clear from this expression that if $\beta<0$, then the above expression is always negative, yielding $\dot\epsilon_1>0$ from Eq.~(\ref{dot-eps1}). 
The constant-roll CI models which are compatible with the observations also restricts the values of $\beta$ in the negative domain \cite{Motohashi:2017aob}. Hence, graceful exit is also ensured in such observationally viable constant-roll CI models. 

However, positive $\beta$ values with 
\begin{eqnarray}
    \beta<\frac{\dot\phi^2}{6M_{\rm Pl}^2H^2}
\end{eqnarray}
can also yield $\dot\epsilon_1>0$, and hence, can end constant-roll CI. Inserting Eq.~(\ref{KG-CR}) into Eq.~(\ref{Hsquare}) and solving for $H^2$ one gets 
\begin{eqnarray}
    H^2&=&\frac{V}{6M_{\rm Pl}^2}\left(1+\sqrt{1+\frac23\frac{M_{\rm Pl}^2}{(1-\beta)^2}\frac{V,_\phi^2}{V^2}}\right)\nonumber\\
    &=&\frac{V}{6M_{\rm Pl}^2}\left(1+\sqrt{1+\frac{4\epsilon_V}{3(1-\beta)^2}}\right),
\end{eqnarray}
which yields 
\begin{eqnarray}
    \beta<\frac{4\epsilon_V}{3(1-\beta)^2}\left(1+\sqrt{1+\frac{4\epsilon_V}{3(1-\beta)^2}}\right)^{-2},
\end{eqnarray}
and further assuming $\beta<1$, the above condition simplifies to $\beta<\epsilon_V/3$. In the above equations, $\epsilon_V$ is a potential slow-roll parameter defined as
$\epsilon_V \equiv (M_{\rm Pl}^2/2)(V,_{\phi}/V)^2.$
Thus one can afford to have positive $\beta$ which ends constant-roll CI, if $\beta$ satisfies the above conditions. To note, this condition clearly depends on the form of the potential. However, observationally, $\beta>0$ models are not favoured in CI because such models yield anomalous super-Hubble evolution of curvature perturbations \cite{Motohashi:2014ppa} \footnote{Note that the constant-roll parameter $\alpha$ defined in \cite{Motohashi:2014ppa} is related to our constant-roll parameter $\beta$ as $\beta=1+\alpha/3$.}. 

\section{Graceful exit condition for constant-roll Warm Inflation}
\label{CR-WI-graceful-exit}

In Warm Inflation (WI), the inflaton field dissipates its energy to a subdominant, yet non-negligible radiation bath throughout inflation. This feature not only modifies the inflaton dynamics, but also affects the background evolution. The Friedmann equations, the Klein-Gordon equation and the evolution of the energy density of the subdominant radiation bath $(\rho_r)$ in WI can be written as
\begin{eqnarray}
&&3M_{\rm Pl}^2H^2=\frac{\dot\phi^2}{2}+V(\phi)+\rho_r,\label{Hsquare-w}\\
&&2M_{\rm Pl}^2\dot H=-\left(\dot\phi^2+\frac43 \rho_r\right),\label{Hdot-w}\\
&&\ddot\phi+3H(1+Q)\dot\phi+V,_\phi=0\label{KG-eq-w}, \\
&&\dot\rho_r+4H\rho_r=\Upsilon\dot\phi^2, \label{rad-eq}
\end{eqnarray}
where $\Upsilon$ designates the rate at which the inflaton dissipates its energy to the radiation bath, and the parameter $Q$ in the Klein-Gordon equation is the ratio of the two frictional terms present in the inflaton dynamics, the friction due to dissipation and the friction due to the expansion of the background universe, and is defined as $Q=\Upsilon/3H$. In general, the dissipative term $\Upsilon$ is a function of the inflaton field ($\phi$) and the temperature of the radiation bath ($T$), assuming the system evolves in a near thermal-equilibrium condition.

For constant-roll WI, we again impose the constant-roll condition, given in Eq.~(\ref{constant-roll-condition}), on the Warm inflationary inflaton dynamics. We need a growing $\epsilon_1$ in order to gracefully exit WI, and from Eq.~(\ref{dot-eps1}) we see that this condition can be achieved in two ways:
\begin{enumerate}
    \item 
$\ddot H<0$ is a sufficient condition to achieve the condition. Given the above set of equations, one gets 
\begin{eqnarray}
M_{\rm Pl}^2\ddot H&=&-\left(-3H\beta\dot\phi^2+\frac23\dot\rho_r\right)\nonumber\\
&=&3H\left(\beta-\frac23 Q\right)\dot\phi^2+\frac83H\rho_r.
\end{eqnarray}
Therefore, the graceful exit condition for constant-roll WI boils down to having 
\begin{eqnarray}
\beta<\frac23 Q-\frac89\frac{\rho_r}{\dot\phi^2}.
\end{eqnarray}
Therefore, as $Q$ is always positive, it is evident that not only negative values of $\beta$, but also positive values of $\beta$ are allowed in WI as far as graceful exit is concerned. However, as the rhs of the above inequality evolves with time, this condition should be maintained throughout WI, and needs to be checked numerically. 
\item If $\ddot H>0$ but $\ddot H<2 \dot H^2/H$, then also one gets growing $\epsilon_1$. However, as this condition doesn't lead to any simplified condition, we leave it at this form. This condition needs to be fulfilled throughout inflation. 
\end{enumerate}


\section{Maintaining thermal equilibrium during constant-roll Warm Inflation}
\label{CR-WI-thermal-equilibrum}

During standard slow-roll WI a constant radiation bath is maintained throughout inflation by the dissipation mechanism. In such a case, one can assume $\dot\rho_r\approx 0$ in Eq.~(\ref{rad-eq}). This radiation bath is also assumed be to evolving near thermal equilibrium, and hence a temperature $T$ can be associated with this radiation bath (the inflaton field, on the other hand, may or may not thermalize with this radiation bath). In such a case, the radiation energy density can be written as 
\begin{eqnarray}
\rho_r=\frac{\pi^2}{30}g_*T^4,\label{rho-T}
\end{eqnarray}
where $g_*$ is the relativistic degrees of freedom of the radiation bath. Furthermore, $\dot\rho_r\approx 0$ also suggests $\dot T\approx 0$ implying that the radiation bath evolves near thermal equilibrium. These features of slow-roll WI suggest that to maintain thermal equilibrium during constant-roll WI  $\dot\rho_r\approx 0$ is a valid assumption. Hence, we will proceed with this condition to analyze the thermal equilibrium during constant-roll WI. With this assumption Eq.~(\ref{rad-eq}) can be written as 
\begin{eqnarray}
\rho_r\approx\frac34 Q\dot\phi^2,\label{rho-approx}
\end{eqnarray}
 and plugging this information into Eq.~(\ref{Hsquare-w}) and Eq.~(\ref{Hdot-w}) yields 
\begin{eqnarray}
&&3M_{\rm Pl}^2H^2\approx\left(1+\frac32 Q\right)\frac{\dot\phi^2}{2}+V(\phi),\label{Hsquare-w-1}\\
&&2M_{\rm Pl}^2\dot H\approx-(1+Q)\dot\phi^2.\label{Hdot-w-1}
\end{eqnarray}
We will proceed with these approximated expressions to determine the conditions for thermal stabilization during constant-roll WI. 

\subsection{The constant $Q$ case with $\beta>0$}
\label{constQ-betap}

We will first assume that $\beta$ is positive. Integrating Eq.~(\ref{constant-roll-condition}) with respect to $N$, we see that 
\begin{eqnarray}
\dot\phi(N)=\dot\phi_0e^{-3\beta N},\label{phi-dot-N}
\end{eqnarray}
where $\dot\phi_0$ is the initial value of $\dot\phi$. In such a case, as inflation progresses the kinetic energy of the inflaton field keeps decreasing. 
Integrating Eq.~(\ref{Hdot-w-1}), with $\dot\phi$ as given above, we get 
\begin{eqnarray}
3M_{\rm Pl}^2H^2=\frac{(1+Q)}{2\beta}\dot\phi_0^2e^{-6\beta N}+V_0, \label{Hsquare-int}
\end{eqnarray}
where $V_0$ is the integration constant. This constant can be determined from the fact that $\epsilon_1$ becomes 1 at the end of inflation ($N=N_f$, say). Therefore, using Eq.~(\ref{Hdot-w-1}) and Eq.~(\ref{Hsquare-int}) to construct $\epsilon_1$ and then setting it to unity at the end of inflation yields 
\begin{eqnarray}
V_0=\left(\frac{3\beta-1}{2\beta}\right)(1+Q)\dot\phi_0^2e^{-6\beta N_f}. \label{V0}
\end{eqnarray}
We note here that $\beta>1/3$ yields positive $V_0$ whereas $\beta<1/3$ yields negative $V_0$. Moreover, to be consistent, Eq.~(\ref{Hsquare-w-1}) and Eq.~(\ref{Hsquare-int}) should be the same which can only happen for a specific form of the potential $V(\phi)$, such as 
\begin{eqnarray}
V(\phi)=\left(\frac{1+Q}{\beta}-1-\frac32Q\right)\frac{\dot\phi_0^2}{2}e^{-6\beta N}+V_0. \label{pot-constQ-betap}
\end{eqnarray}
We will later determine the form of the potential in terms of $\phi$.

We will now determine what values of $\beta$ can yield temperature stabilization during constant-roll WI. Assuming a constant radiation bath is maintained during constant-roll WI, we determine the evolution of the temperature $T$ of the radiation bath as
\begin{eqnarray}
\frac{d\ln T}{dN}=\frac{1}{4-p}\left(\frac{\Upsilon,_\phi}{H\Upsilon}\dot\phi+\epsilon_1-6(1+Q)-2\frac{V,_\phi}{H\dot\phi}\right), \label{dlntdN}
\end{eqnarray}
where $p\equiv (\Upsilon,_TT)/\Upsilon$. We have used Eq.~(\ref{rho-T}), Eq.~(\ref{rho-approx}) and Eq.~(\ref{KG-eq-w}) to determine the above expression. For constant $Q$ we see that 
\begin{eqnarray}
\frac{\Upsilon,_\phi}{H\Upsilon}\dot\phi=\frac{1}{3QH^2}\frac{d}{dt}(3QH)\frac{dt}{d\phi}\dot\phi=-\epsilon_1.
\end{eqnarray}
In addition to that, the potential given in Eq.~(\ref{pot-constQ-betap}) yields
\begin{eqnarray}
V,_\phi=3H\dot\phi\left[\beta\left(1+\frac32Q\right)-(1+Q)\right]. \label{V-phi-constQ-betap}
\end{eqnarray}
Plugging the above two expressions in Eq.~(\ref{dlntdN}), we get 
\begin{eqnarray}
\frac{d\ln T}{dN}=-\frac{6\beta}{4-p}\left(1+\frac32Q\right).
\end{eqnarray}
Now, if WI is taking place in a weak dissipative regime $(Q\ll1)$ then one requires $\beta\ll1$ to maintain thermal equilibrium during constant-roll. However, if WI is taking place in strong dissipative require $(Q\gg1)$, then thermal stabilization during constant-roll demands $Q\beta\ll1$. In both the scenarios $V_0$ will become negative because one requires $\beta\ll1$, as can be seen from Eq.~(\ref{V0}).

Furthermore, we note that plugging the constant-roll condition Eq.~(\ref{constant-roll-condition}) in the WI inflaton equation of motion, given in Eq.~(\ref{KG-eq-w}), one obtains
\begin{eqnarray}
V,_\phi=3H\dot\phi[\beta-(1+Q)]. \label{KG-Warm-CR}
\end{eqnarray}
It is required that the potential given in Eq.~(\ref{pot-constQ-betap}) should yield the same equation of motion of the inflaton field during constant roll. Hence, Eq.~(\ref{V-phi-constQ-betap}) should resemble the above equation. In the weak dissipative regime, if $Q\ll2/3$, then these two expressions are equivalent. This imposes another condition on weak dissipative constant-roll WI models with constant $Q$. However, in the strong dissipative regime, these two expressions become equivalent when $\beta\ll 2/3$, which is consistent with the thermal stabilization condition. 

To summarize, thermal stabilization can be maintained during constant-roll WI (in models where $Q$ is constant) if
\begin{itemize}
\item in weak dissipative regime $(Q\ll1)$ one has $\beta\ll1$ and $Q\ll2/3$
\item in strong dissipative regime $(Q\gg1)$ one has $Q\beta\ll1$ and $\beta\ll2/3$.
\end{itemize}

\begin{figure}
    \centering
    \includegraphics[width=8cm]{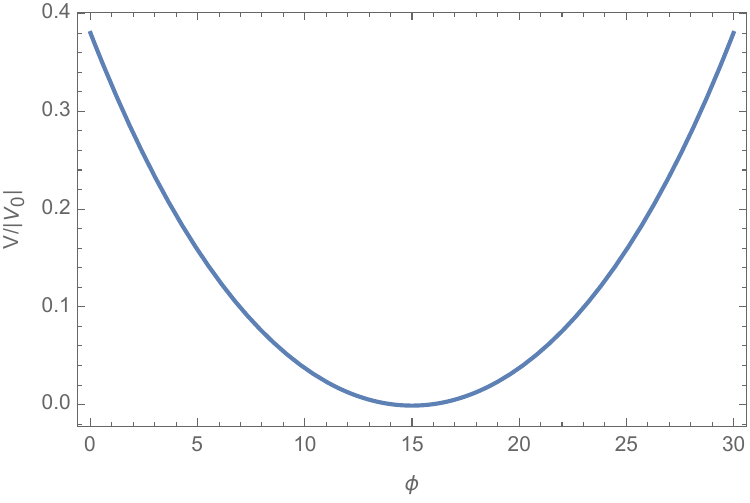}
    \caption{The form of the potential given in Eq.~(\ref{pot-1}), with $\beta=10^{-3}$, $Q=10^{-2}$ and $\phi_0=15$.}
    \label{pot-beta-positive-Q-const}
\end{figure}

We will now determine the form of the potential, given in Eq.~(\ref{pot-constQ-betap}), in terms of $\phi$ which gives rise to constant-roll WI. Integrating Eq.~(\ref{phi-dot-N}) again with respect to $N$ (while keeping in mind that $V_0$ is negative) we get
\begin{eqnarray}
\phi(N)=\phi_0-\frac{M_{\rm Pl}}{3}\sqrt{\frac{6}{\beta(1+Q)}}\cosh^{-1}\left(\sqrt{\frac{1+Q}{2\beta |V_0|}}\dot\phi_0e^{-3\beta N}\right)\,,\label{phi-N}\nonumber\\
\end{eqnarray}
where $|V_0|$ is the absolute value of $V_0$ and $\phi_0$ is the integration constant. Here $\phi_0$ may not be equal to $\phi(N=0)$. 
Rearranging Eq.~(\ref{phi-N}) we find
\begin{eqnarray}
\dot\phi_0e^{-3\beta N}=\sqrt{\frac{2\beta |V_0|}{1+Q}}\cosh\left(\frac{3}{M_{\rm Pl}}\sqrt{\frac{\beta(1+Q)}{6}}(\phi_0-\phi)\right). \label{phi0-sol}\nonumber\\
\end{eqnarray}
Plugging this expression into Eq.~(\ref{pot-constQ-betap}) we find the form of the potential giving rise to constant-roll WI as 
\begin{widetext}
\begin{eqnarray}
V(\phi)=|V_0|\left[\left(1-\frac{\beta(2-3Q)}{2(1+Q)}\right)\cosh^2\left(\frac{3}{M_{\rm Pl}}\sqrt{\frac{\beta(1+Q)}{6}}(\phi_0-\phi)\right)-1\right].
\label{pot-1}
\end{eqnarray}
\end{widetext}
Fig~(\ref{pot-beta-positive-Q-const}) depicts the form of this potential. 
This potential corresponds to the Hubble parameter 
\begin{eqnarray}
H=\sqrt{\frac{|V_0|}{3M_{\rm Pl}^2}}\sinh\left(\sqrt{\frac{3\beta(1+Q)}{2}}\frac{(\phi-\phi_0)}{M_{\rm Pl}}\right).
\end{eqnarray}
The above expression can be obtained by substituting Eq.~(\ref{phi0-sol}) into Eq.~(\ref{Hsquare-int}).
One can tally these results with the constant-roll Cold Inflation case by setting $Q=0$ (and replacing $\beta$ with $1+\alpha/3$ to compare with \cite{Motohashi:2014ppa}), which yields the Hubble parameter given in Eq.~(2.13) of \cite{Motohashi:2014ppa} and its corresponding potential given in Eq.~(2.26). However, unlike the constant-roll WI case we are discussing here, the corresponding constant-roll CI case doesn't give rise to an inflationary phase because it yields $\ddot{a}(t)<0$, as has been pointed out in \cite{Motohashi:2014ppa}. \footnote{The constant-roll WI scenarios corresponding to the other two constant-roll CI scenarios discussed in \cite{Motohashi:2014ppa} with Hubble parameters given in Eq.~(2.11) and Eq.~(2.13) along with their corresponding potentials, given in Eq.~(2.14) and Eq.~(2.18), can be found by considering $V_0=0$ and $V_0>0$ in Eq.~(\ref{Hsquare-int}), respectively, and finding their corresponding potentials. However, these two constant-roll WI scenarios cannot be realised while maintaining near thermal equilibrium during inflation.}

\subsubsection{Numerical analysis of constant-roll WI models in weak dissipative regime}

We evolve a constant-roll Warm Inflation scenario, with the dynamical equations given in Eq.~(\ref{Hsquare-w}), Eq.~(\ref{KG-Warm-CR})and Eq.~(\ref{rad-eq}), fully numerically with the choice of parameter values as $\beta=10^{-3}$, $Q=10^{-2}$, $g_*=106.75$, $\phi_0=15$ and initial conditions as $T(N=0)=2.79\times 10^{-6}\, M_{\rm Pl}$ and $\dot\phi(N=0)\equiv \dot\phi_0=5.33\times10^{-10}\, M_{\rm Pl}^2$. Once the parameters are fixed we also know the values of $V_0$ from Eq.~(\ref{V0}) by setting $N_f=60$. 
We get, with the above choice of parameters, $|V_0|\approx10^{-16}\, M_{\rm Pl}^4$ and $\phi(N=0)=-1\,M_{\rm Pl}$. We also note that our choice of parameters are such that $\beta\ll1$ and $Q\ll2/3$, which are the theoretically determined conditions to be met to have a constant-roll WI in weak dissipative regime with constant $Q$ models which can evolve near thermal equilibrium throughout inflation. 

\begin{center}
\begin{figure*}[!htb]
\subfigure[Graceful exit condition]{\includegraphics[width=6cm]{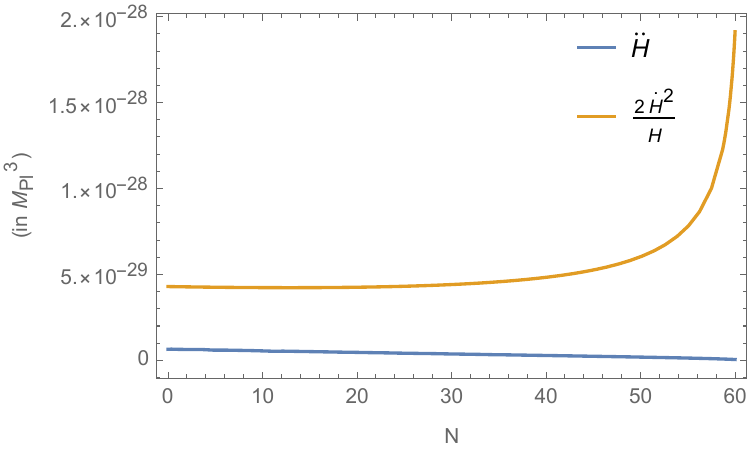}}
\subfigure[Evolution of $\epsilon_1$]{\includegraphics[width=5.2cm]{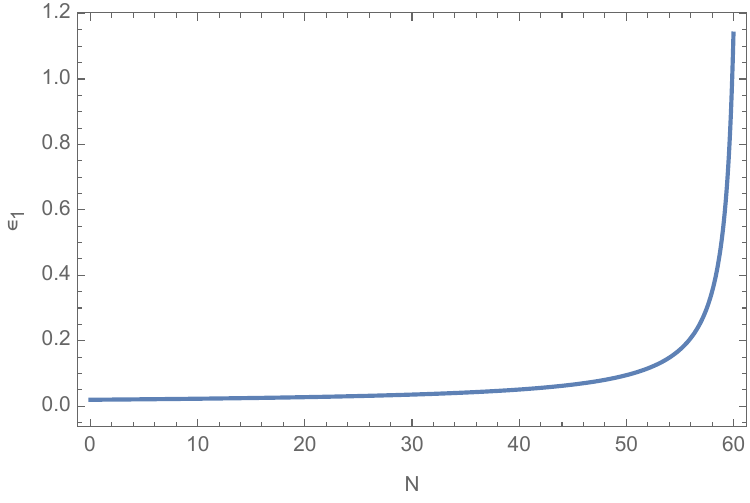}}
\subfigure[Evolution of $\epsilon_2$]{\includegraphics[width=5.2cm]{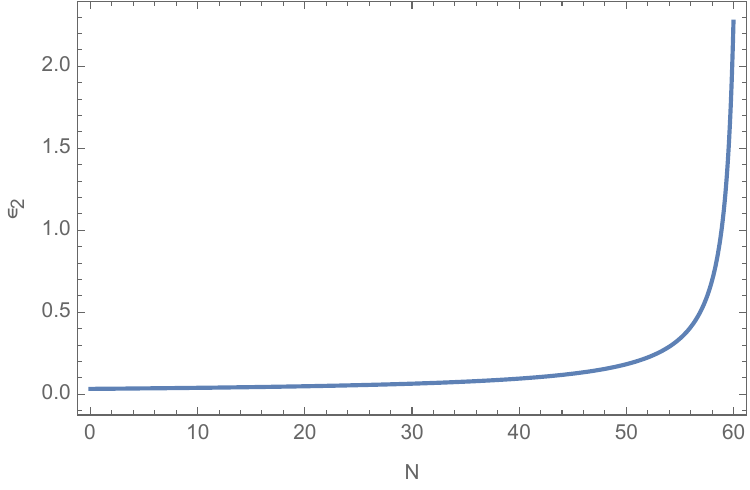}}\\
\subfigure[Evolution of temperature $T$]{\includegraphics[width=6cm]{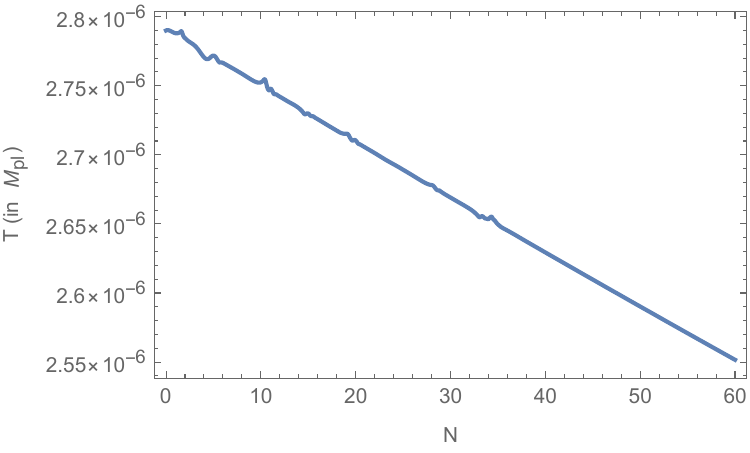}}
\subfigure[Thermalization condition $T>H$]{\includegraphics[width=6cm]{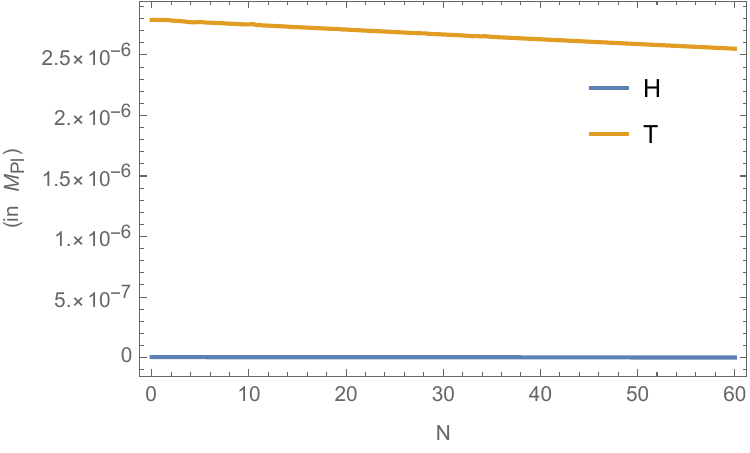}}
\caption{Results of numerical evolution of a constant-roll WI scenario with positive $\beta$ and constant $Q$ in weak dissipative regime.}
\label{weak-const-Q}
\end{figure*}
\end{center}

The results of this numerical analysis is depicted in Fig.~(\ref{weak-const-Q}). As in this model $\ddot H>0$ throughout, thus to have $\dot\epsilon_1>0$ one needs to maintain $\ddot H< 2\dot H^2/H$ throughout inflation in order to gracefully exit inflation. Sub-figure (a) confirms that this condition has been maintained. Sub-figures (b) and (c) show the evolution of $\epsilon_1$ and $\epsilon_2$ of this model, and we see that both the slow-roll parameters remain less than unity throughout the evolution. The temperature of the radiation bath doesn't evolve much during the inflationary phase as has been shown in sub-figure (d), along with the thermalization condition $T>H$ maintained throughout, indicating near thermal evolution of the system. 

\subsubsection{Numerical analysis of constant-roll WI models in strong dissipative regime}

We again evolve Eq.~(\ref{Hsquare-w}), Eq.~(\ref{KG-Warm-CR})and Eq.~(\ref{rad-eq}) fully numerically, now with the choice of parameter values as $\beta=10^{-4}$, $Q=2.8\times10^{2}$, $g_*=106.75$, $\phi_0=0$ and initial conditions as $T(N=0)=0.6\times 10^{-6}\, M_{\rm Pl}$ and $\dot\phi(N=0)\equiv \dot\phi_0=8.59\times10^{-14}\, M_{\rm Pl}^2$. Once the parameters are fixed we get $|V_0|\approx10^{-20}\, M_{\rm Pl}^4$ and $\phi(N=0)=-1\,M_{\rm Pl}$. We also note that our choice of parameters are such that $\beta\ll2/3$ and $Q\beta\ll1$, which are the theoretically determined conditions to be met to have a constant-roll WI in strong dissipative regime with constant $Q$ models which can evolve near thermal equilibrium throughout inflation. 

\begin{center}
\begin{figure*}[!htb]
\subfigure[Graceful exit condition]{\includegraphics[width=6cm]{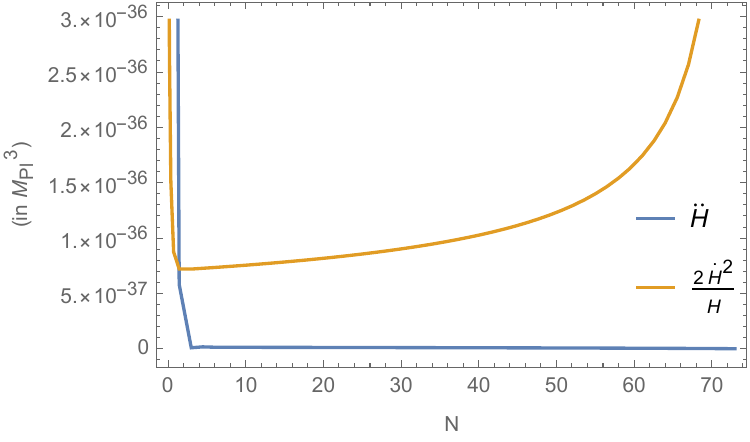}}
\subfigure[Evolution of $\epsilon_1$]{\includegraphics[width=5.2cm]{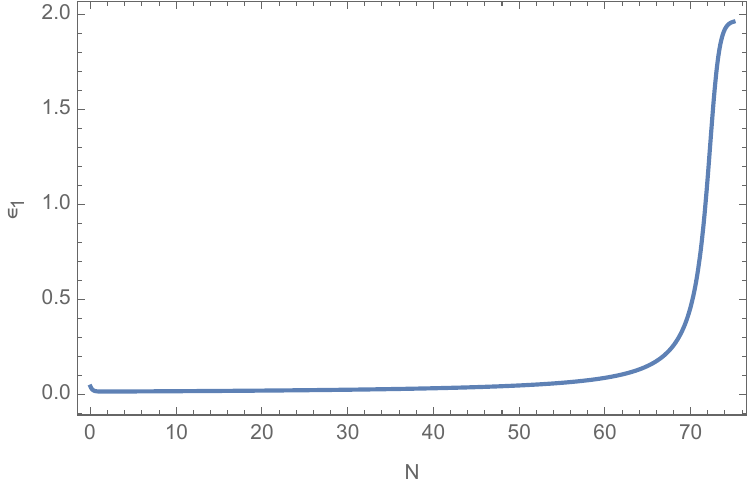}}
\subfigure[Evolution of $\epsilon_2$]{\includegraphics[width=5.2cm]{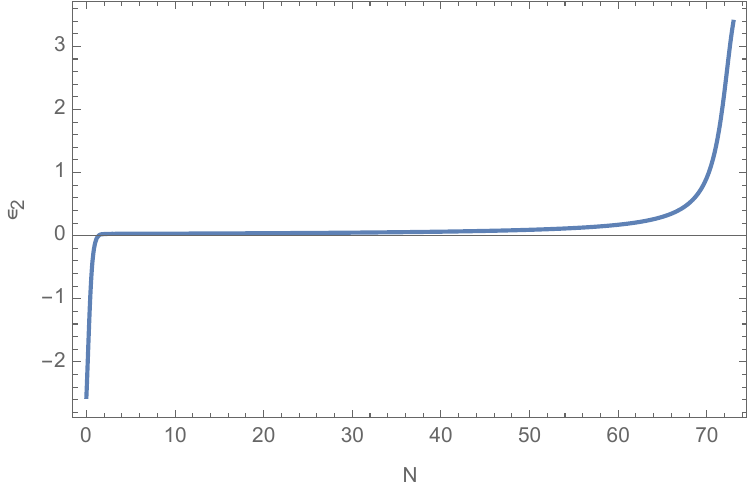}}\\
\subfigure[Evolution of temperature $T$]{\includegraphics[width=6cm]{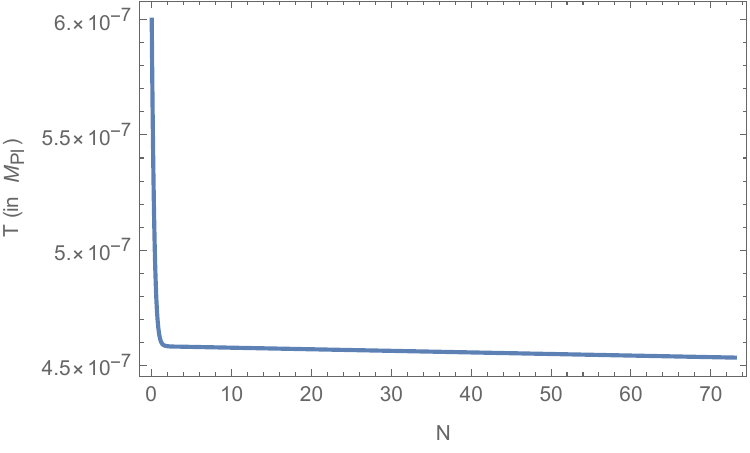}}
\subfigure[Thermalization condition $T>H$]{\includegraphics[width=6cm]{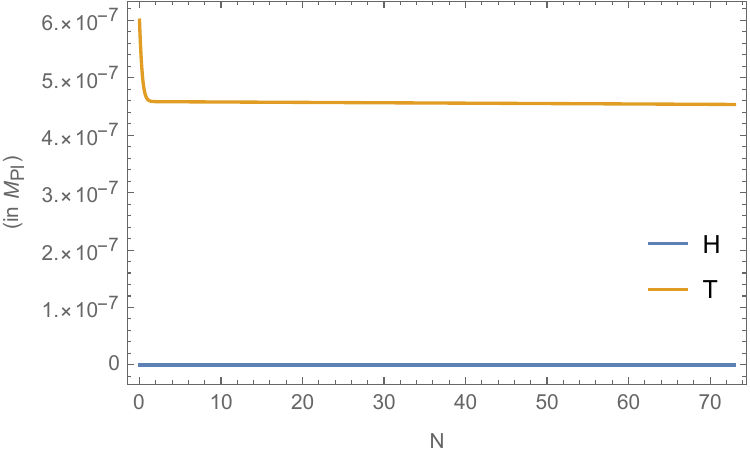}}
\caption{Results of numerical evolution of a constant-roll WI scenario with positive $\beta$ and constant $Q$ in strong dissipative regime.}
\label{strong-const-Q}
\end{figure*}
\end{center}

The results of this numerical analysis is depicted in Fig.~(\ref{strong-const-Q}). As in this model too $\ddot H$ remains positive throughout, sub-figure (a) confirms that the condition, $\ddot H< 2\dot H^2/H$, has been maintained throughout and thus guarantees graceful exit. Sub-figures (b) and (c) show the evolution of $\epsilon_1$ and $\epsilon_2$ of this model, and we see that though $\epsilon_1$ remains less than unity throughout the evolution, $\epsilon_2$ starts with a value larger than unity, but quickly becomes less than unity and remains so for the rest of the evolution till inflation lasts. The temperature of the radiation bath doesn't evolve much during the inflationary phase as has been shown in sub-figure (d). The thermalization condition $T>H$ maintained throughout, indicating near thermal evolution of the system. 

\subsubsection{Some comments on the analysis presented in Ref.~\cite{Kamali:2019wdh} and Ref.~\cite{Mun:2021kzb}}

The first paper which made an attempt to formulate constant-roll WI is \cite{Kamali:2019wdh}. The authors in this paper formulated the background evolution of the inflationary universe undergoing constant-roll WI and presented the results related to the cosmological perturbations during such a constant-roll WI. A few features of this analysis which we think are required to be reconsidered are as follows:
\begin{enumerate}
    \item The paper doesn't specify whether their model is suitable for sustaining at least 60 $e-$folds of constant-roll inflation or whether it is related to a transient constant-roll phase. This issue stems from the fact the authors did not comment on graceful exit from a constant-roll phase. 
    \item The analysis is done for positive $\beta$ values only. Also, the weak dissipative regime is not explored in the paper. As we have shown here, constant-roll WI can be realized in both weak and strong dissipative regimes with both positive and negative values of $\beta$. 
    \item The analysis is done only for toy models of WI where $Q$ is treated as constant. In realistic models of WI, $Q$ becomes a function of both $\phi$ and $T$. 
    \item The background analysis done in this paper indicates that the dynamics doesn't evolve in a near thermal equilibrium condition. Yet, when the perturbation calculations are done, it appears that the temperature parameter $T$ is taken to be a constant. The references used by the previous authors to write the Langevin equation, (Eq.~(4.5) in Ref.~\cite{Kamali:2019wdh}) explicitly requires the system to be have a nearly constant temperature $T$. Thus the background analysis doesn't provide the environment for the perturbation analysis presented in the paper. 
    \item The authors mention that the constant-roll WI dynamics appears to be an interplay between both cold and warm inflation. The challenge one might face while considering CI and WI simultaneously is that the perturbation analysis in both these scenarios are vastly different. In CI, the generated perturbations determining the scalar power spectrum are quantum in nature, while in WI they are classical being thermal fluctuations. Thus, treating CI and WI simultaneously, if at all called for, requires more serious consideration. 
\end{enumerate}

The second paper ~\cite{Mun:2021kzb} which also addresses the issue of constant-roll WI defines the constant-roll condition for WI in a rather different way:
\begin{eqnarray}
\ddot{\phi}=-3H(1+Q)\lambda \dot{\phi},
\end{eqnarray}
where $\lambda$ is a dimensionless constant. This stems form the fact that in the inflaton's equation of motion of WI, not only the Hubble friction term has a $\dot\phi$ factor, but the dissipative term also has one. This raises the question whether the constant rate of roll of the inflaton field in a constant-roll WI scenario should depend on both these terms in the inflaton's equation of motion. We think that as $Q$ is not a constant in realistic models of WI, the above way of defining the constant-roll condition would not lead to constant rate of roll of the inflaton field in WI. Thus we stick to the original constraint (Eq.~(\ref{constant-roll-condition})) as has been defined in cases of CI. The analysis presented in this paper, too, does not address the issues of graceful exit or thermalization condition (thermalization is assumed, without showing whether the thermalization condition can be maintained throughout). Above all the parametrization of $Q$ as a function of $\phi$ and $T$ considered in this paper are rather ad hoc and doesn't stem from any realistic model of WI. This point has already been mentioned in the introduction of the paper.


\subsection{The constant $Q$ case with $\beta<0$}

We will do a similar analysis as above, but now with a negative $\beta$. We define $\tilde\beta=|\beta|$. We first note that with the constant-roll constraint equation, Eq.~(\ref{constant-roll-condition}), $\dot\phi$ will now be an increasing function of $e-$folds:
\begin{eqnarray}
\dot\phi(N)=\dot\phi_0e^{3\tilde\beta N},\label{phi-dot-N-neg-beta}
\end{eqnarray}
which then yields using Eq.~(\ref{Hdot-w-1})
\begin{eqnarray}
3M_{\rm Pl}^2H^2=-\frac{(1+Q)}{2\tilde\beta}\dot\phi_0^2e^{6\tilde\beta N}+V_0. \label{Hsquare-int-neg-beta}
\end{eqnarray}
It is to note here that, unlike the previous case with $\beta>0$, here $V_0$ has to be positive as the first term on rhs of the above equation is always negative throughout inflation, and it is required that $V_0$ always remains greater than the first term so that $H$ remains real. 
One can determine $V_0$ as before by setting $\epsilon_1=1$ at $N=N_f$ and get
\begin{eqnarray}
V_0=\left(\frac{1+3\tilde\beta}{2\tilde\beta}\right)(1+Q)\dot\phi_0^2e^{6\tilde\beta N_f}, \label{V0-betan}
\end{eqnarray}
which turns out to be positive. Also, the form of the potential which is consistent with both Eq.~(\ref{Hsquare-w-1}) and Eq.~(\ref{Hsquare-int-neg-beta}) is 
\begin{eqnarray}
V(\phi)=V_0-\left(1+\frac32Q+\frac{1+Q}{\beta}\right)\frac{\dot\phi_0^2}{2}e^{6\tilde\beta N}. \label{pot-constQ-betan}
\end{eqnarray}

Proceeding with the same analysis as before, we find the temperature evolution in such cases can be written as 
\begin{eqnarray}
\frac{d\ln T}{dN}=\frac{6\tilde\beta}{4-p}\left(1+\frac32Q\right).
\end{eqnarray}
Therefore, as before, temperature stabilizes in weak dissipative regime $(Q\ll1)$ when $\tilde\beta\ll1$ and in the strong dissipative regime $(Q\gg1)$ when $Q\tilde\beta\ll1$. Also, comparing the Klein-Gordon equation of WI including the constant-roll constraint,
\begin{eqnarray}
V,_\phi=-3H\dot\phi[\tilde\beta+(1+Q)], \label{KG-Warm-CR-betan}
\end{eqnarray}
with the derivative of the potential obtained in Eq.~(\ref{pot-constQ-betan}) as
\begin{eqnarray}
V,_\phi=-3H\dot\phi\left[\tilde\beta\left(1+\frac32Q\right)+(1+Q)\right], \label{V-phi-constQ-betan}
\end{eqnarray}
we see that one also requires $Q\ll 2/3$ in the weak dissipative regime and $\tilde\beta\ll2/3$ in the strong dissipative regime to have a consistent picture. Thus, we obtained similar conditions for near thermal equilibrium as before in the weak and strong dissipative regime. 

\begin{figure}
    \centering
    \includegraphics[width=8cm]{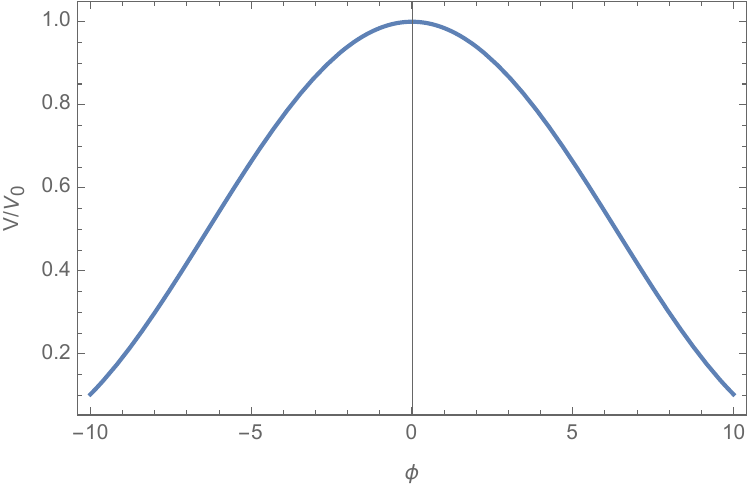}
    \caption{The form of the potential given in Eq.~(\ref{pot-2}), with $\beta=-10^{-2}$, $Q=10^{-2}$ and $\phi_0=0$.}
    \label{pot-beta-negative-Q-const}
\end{figure}

Finally to determine the form of the potential as a function of $\phi$ we proceed as before: first integrate Eq.~(\ref{phi-dot-N-neg-beta}) to obtain $\phi(N)$ as 
\begin{eqnarray}
\phi(N)=\frac{M_{\rm Pl}}{3}\sqrt{\frac{6}{\tilde\beta(1+Q)}}\sin^{-1}\left(\sqrt{\frac{1+Q}{2V_0\tilde\beta}}\dot\phi_0e^{3\tilde\beta N}\right)-\phi_0, \nonumber\\ \label{phiN-betan}
\end{eqnarray}
rearranging the expression to obtain $\dot\phi_0e^{3\tilde\beta N}$, and then insert it into Eq.~(\ref{pot-constQ-betan}). We get 
\begin{widetext}
\begin{eqnarray}
V(\phi)=V_0\left[1-\left(1+\frac{\tilde\beta(2+3Q)}{2(1+Q)}\right)\sin^2\left(\frac{3}{M_{\rm Pl}}\sqrt{\frac{\tilde\beta(1+Q)}{6}}(\phi_0+\phi)\right)\right].
\label{pot-2}
\end{eqnarray}
\end{widetext}
Fig~(\ref{pot-beta-negative-Q-const}) depicts the form of this potential. 
This potential corresponds to the Hubble parameter 
\begin{eqnarray}
H=\sqrt{\frac{V_0}{3M_{\rm Pl}^2}}\cos\left(\sqrt{\frac{3\tilde\beta(1+Q)}{2}}\frac{(\phi+\phi_0)}{M_{\rm Pl}}\right).
\end{eqnarray}
As $V_0>0$ in this case, it corresponds to the Hubble parameter of constant-roll CI as given in Eq.~(2.12) of \cite{Motohashi:2014ppa} and the above potential corresponds to the potential given in Eq.~(2.22) of \cite{Motohashi:2014ppa} in the $Q\rightarrow 0$ limit.

\subsubsection{Numerical analysis of constant-roll WI models in weak dissipative regime}

We evolve Eq.~(\ref{Hsquare-w}), Eq.~(\ref{KG-Warm-CR-betan})and Eq.~(\ref{rad-eq}) fully numerically, with the choice of parameter values as $\beta=-10^{-2}$, $Q=10^{-2}$, $g_*=106.75$, $\phi_0=0$ and initial conditions as $T(N=0)=1.56\times 10^{-6}\, M_{\rm Pl}$ and $\dot\phi(N=0)\equiv \dot\phi_0=1.7\times10^{-10}\, M_{\rm Pl}^2$. Once the parameters are fixed we get $V_0\approx10^{-16}\, M_{\rm Pl}^4$ (from Eq.~(\ref{V0-betan})) and $\phi(N=0)=1\,M_{\rm Pl}$ (from Eq.~(\ref{phiN-betan})). We also note that our choice of parameters are such that $\tilde\beta\ll1$ and $Q\ll2/3$, which are the theoretically determined conditions to be met to have a constant-roll WI in weak dissipative regime with constant $Q$ models to evolve near thermal equilibrium throughout inflation. 

\begin{center}
\begin{figure*}[!htb]
\subfigure[Graceful exit condition]{\includegraphics[width=6cm]{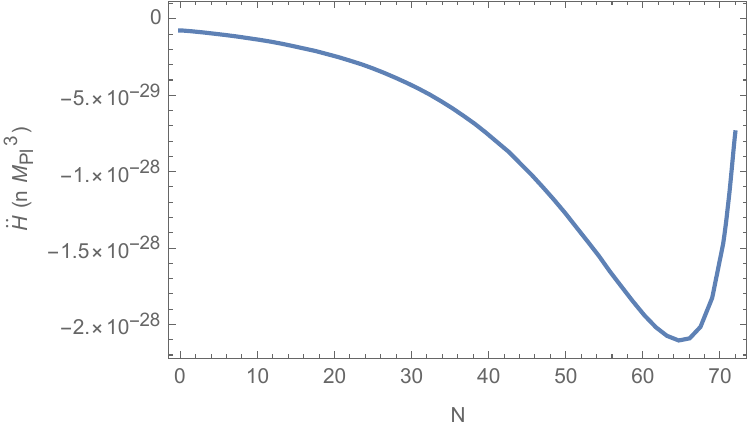}}
\subfigure[Evolution of $\epsilon_1$]{\includegraphics[width=5.2cm]{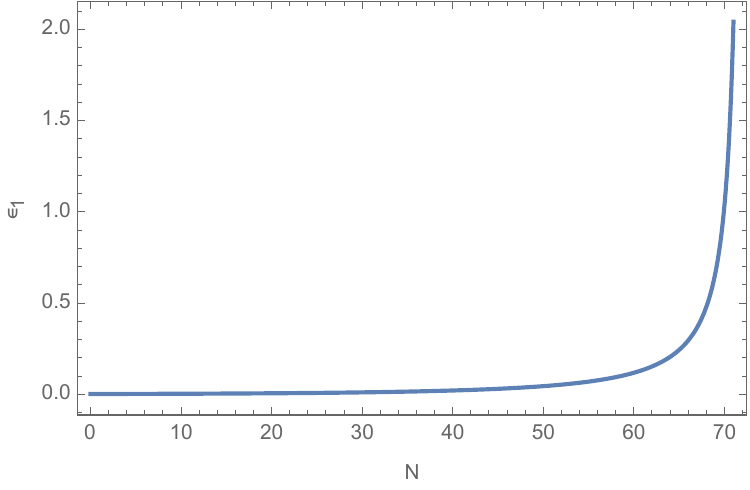}}
\subfigure[Evolution of $\epsilon_2$]{\includegraphics[width=5.2cm]{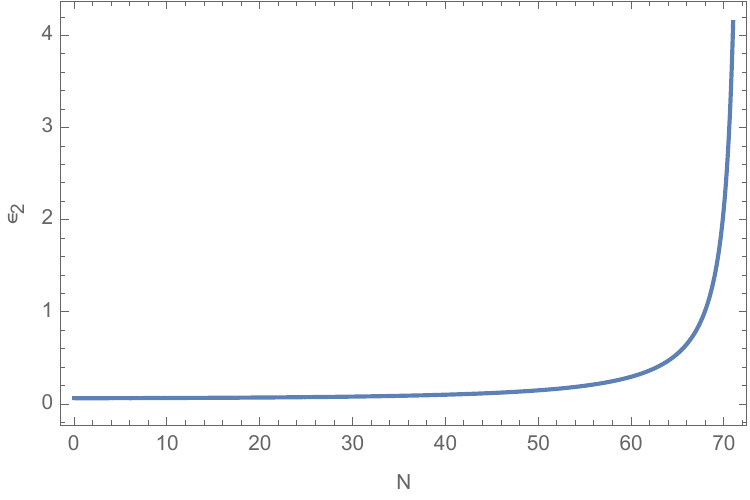}}\\
\subfigure[Evolution of temperature $T$]{\includegraphics[width=6cm]{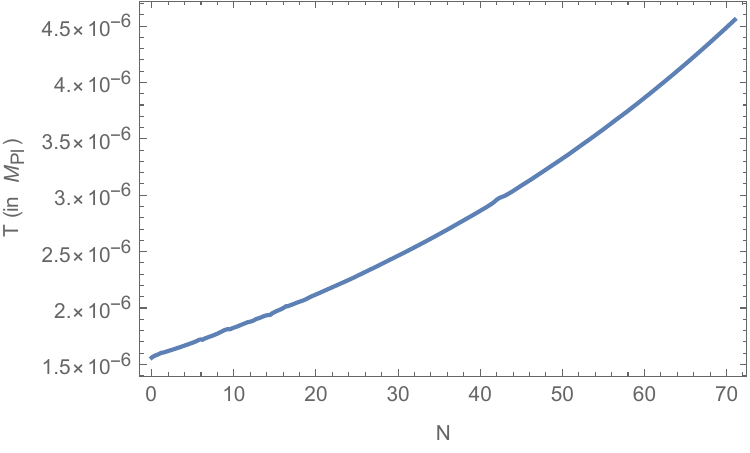}}
\subfigure[Thermalization condition $T>H$]{\includegraphics[width=6cm]{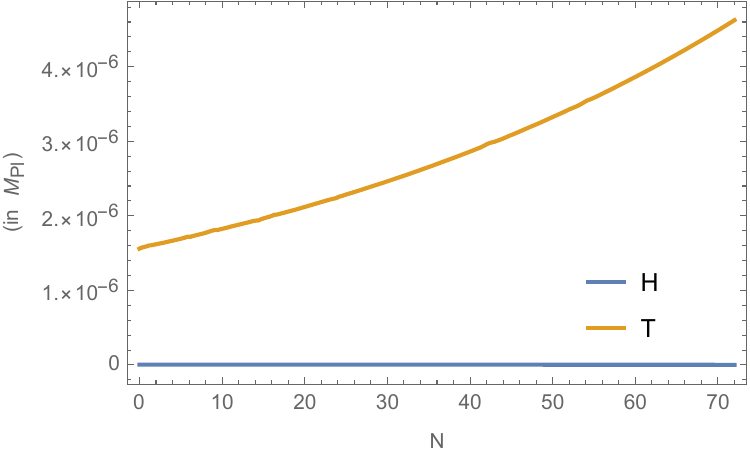}}
\caption{Results of numerical evolution of a constant-roll WI scenario with negative $\beta$ and constant $Q$ in weak dissipative regime.}
\label{weak-const-Q-betan}
\end{figure*}
\end{center}

The results of this numerical analysis is depicted in Fig.~(\ref{weak-const-Q-betan}). As in this model $\ddot H$ remains negative throughout, as can be seen from sub-figure (a), graceful exit is always guaranteed. Sub-figures (b) and (c) show the evolution of $\epsilon_1$ and $\epsilon_2$ of this model, and we see that both $\epsilon_1$ and $\epsilon_2$ remain less than unity throughout the evolution. The temperature of the radiation bath doesn't evolve much during the inflationary phase as has been shown in sub-figure (d), along with the thermalization condition $T>H$ maintained throughout, indicating near thermal evolution of the system. 

\subsubsection{Numerical analysis of constant-roll WI models in strong dissipative regime}

We again evolve Eq.~(\ref{Hsquare-w}), Eq.~(\ref{KG-Warm-CR-betan})and Eq.~(\ref{rad-eq}) fully numerically, now with the choice of parameter values as $\beta=-10^{-4}$, $Q=2.8\times 10^{2}$, $g_*=106.75$, $\phi_0=7.65$ and initial conditions as $T(N=0)=1.56\times 10^{-6}\, M_{\rm Pl}$ and $\dot\phi(N=0)\equiv \dot\phi_0=8.28\times10^{-12}\, M_{\rm Pl}^2$. Once the parameters are fixed we get $V_0\approx10^{-16}\, M_{\rm Pl}^4$ (from Eq.~(\ref{V0-betan})) and $\phi(N=0)=-1\,M_{\rm Pl}$ (from Eq.~(\ref{phiN-betan})). We also note that our choice of parameters are such that $\tilde\beta\ll2/3$ and $Q\tilde\beta\ll1$, which are the theoretically determined conditions to be met to have a constant-roll WI in strong dissipative regime with constant $Q$ models to evolve near thermal equilibrium throughout inflation. 

\begin{center}
\begin{figure*}[!htb]
\subfigure[Graceful exit condition]{\includegraphics[width=6cm]{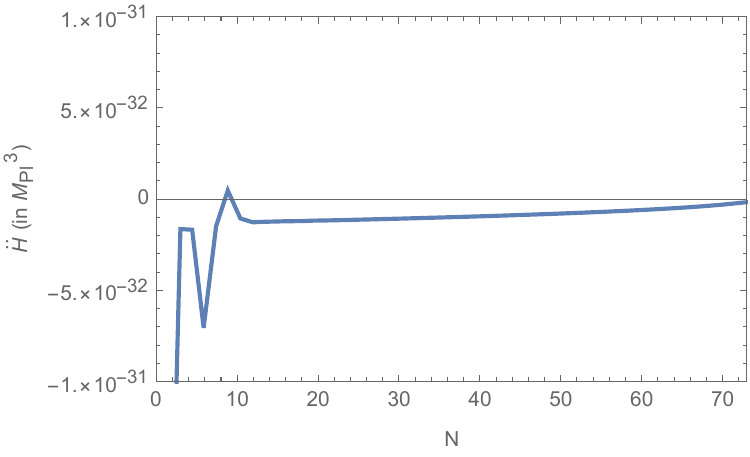}}
\subfigure[Evolution of $\epsilon_1$]{\includegraphics[width=5.2cm]{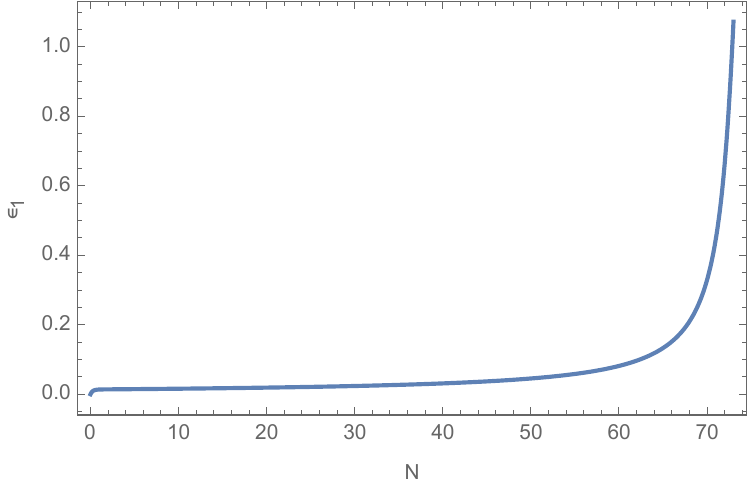}}
\subfigure[Evolution of $\epsilon_2$]{\includegraphics[width=5.2cm]{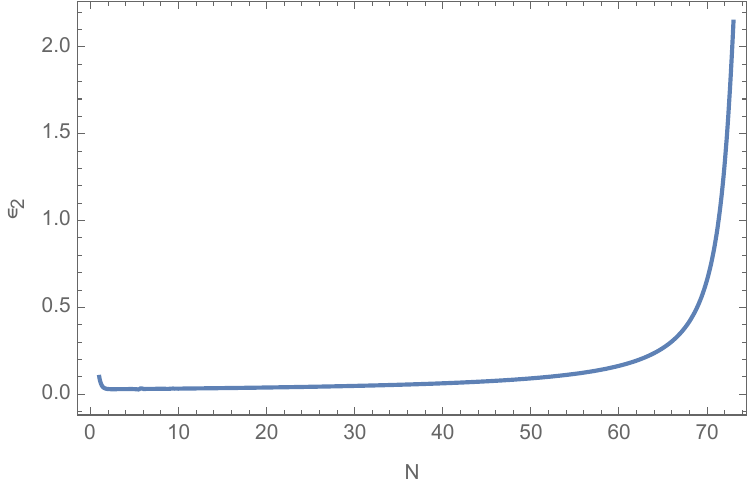}}\\
\subfigure[Evolution of temperature $T$]{\includegraphics[width=6cm]{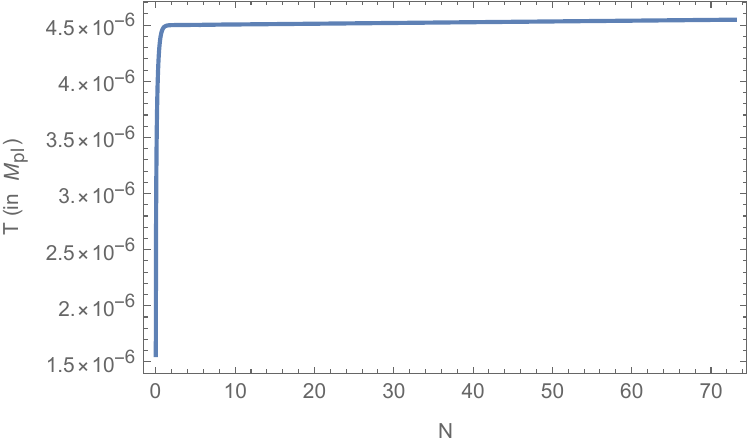}}
\subfigure[Thermalization condition $T>H$]{\includegraphics[width=6cm]{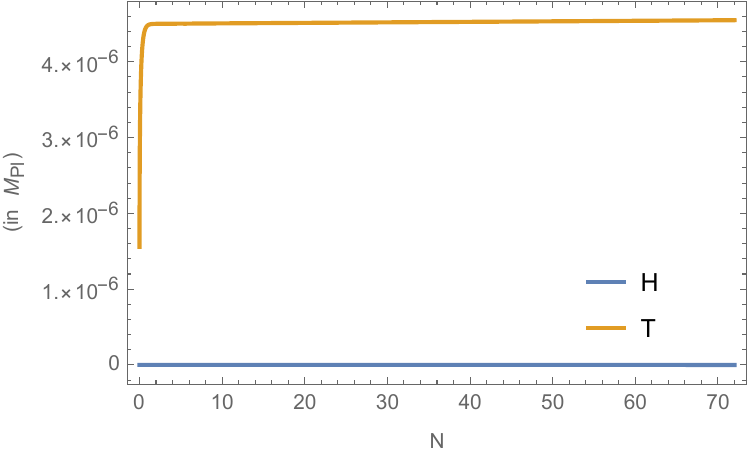}}
\caption{Results of numerical evolution of a constant-roll WI scenario with negative $\beta$ and constant $Q$ in weak dissipative regime.}
\label{strong-const-Q-betan}
\end{figure*}
\end{center}

The results of this numerical analysis is depicted in Fig.~(\ref{strong-const-Q-betan}). Like in the previous case, $\ddot H$ remains negative throughout, as can be seen from sub-figure (a), and thus graceful exit is always guaranteed. Sub-figures (b) and (c) show the evolution of $\epsilon_1$ and $\epsilon_2$ of this model, and we see that both $\epsilon_1$ and $\epsilon_2$ remain less than unity throughout the evolution. The temperature of the radiation bath doesn't evolve much during the inflationary phase as has been shown in sub-figure (d), along with the thermalization condition $T>H$ maintained throughout, indicating near thermal evolution of the system. 

\subsection{The case with $\Upsilon$ as a function of $T$ alone}

This case resembles the analysis of the constant $Q$ case. Let us first discuss a subclass of such models with $\beta>0$. We see from Eq.~(\ref{dlntdN}) that if $\Upsilon$ is a function of $T$ alone, then $\Upsilon,_\phi=0$. In such a case the evolution of temperature with $e-$folds can be written as 
\begin{eqnarray}
\frac{d\ln T}{dN}=\frac{1}{4-p}\left[\epsilon_1-6\beta\left(1+\frac32Q\right)\right].
\end{eqnarray}
As $\epsilon_1$ remains much smaller than unity throughout constant-roll WI, the thermal stabilization condition remains the same as has been pointed out in the case of constant $Q$ in Subsection~\ref{constQ-betap}. As we can see that the rest of the calculations of the constant $Q$ case don't depend on the form of $\Upsilon$, the outcome of the case where $\Upsilon$ is a function of $T$ alone is identical to the constant $Q$ case, but now with $p\neq0$. A similar argument can also be presented for the subclass of such models with negative $\beta$. 

There are several models of Warm Inflation where the dissipative coefficient $\Upsilon$ is a function of $T$ alone. For example, the Minimal Warm Inflation model ($\Upsilon\propto T^3$, and thus $p=3$) \cite{Berghaus:2019whh} and the model presented in \cite{Bastero-Gil:2019gao} ($\Upsilon\propto T^{-1}$, and thus $p=-1$) are models of WI where WI is realized in the strong dissipative regime, whereas the Warm Little inflaton model ($\Upsilon\propto T$, and thus $p=1$) is an example where WI is realized in the weak dissipative regime. All these models can be realised with constant-roll dynamics as well in which the system will evolve in a near thermal equilibrium condition. 


\subsection{The case with $\Upsilon$ as a function of $\phi$}

The situation drastically changes when we deal with WI models where the dissipative coefficient $\Upsilon$ is a function of $\phi$. A general form of the dissipative coefficients generally appearing in WI models can be written as 
\begin{eqnarray}
\Upsilon(\phi,T)=C_\Upsilon T^p\phi^cM^{1-p-c},
\end{eqnarray}
where the power of $T$ and $\phi$, i.e. $p$ and $c$ respectively, can take integer values (positive or negative), $C_\Upsilon$ is a dimensionless quantity which depends on the microscopic details of the WI model under consideration and $M$ is some appropriate mass scale so that the dimensionality of $\Upsilon$ is kept at $[\Upsilon]=[{\rm mass}]$. The stability of WI models restricts the dependance of $T$ of $\Upsilon$ to $|p|\leq 4$ \cite{Moss:2008yb}. 

If we want such models to evolve with constant-roll dynamics in a near thermal equilibrium condition, we can treat $T$ to be a constant. Also, both Eq.~(\ref{rho-T}) and Eq.~(\ref{rho-approx}) should hold. Equating these two equations with the above form of $\Upsilon(\phi,T)$ one gets 
\begin{eqnarray}
H=\frac{C_\Upsilon M^{1-p-c}}{4C_r}T^{p-4}\phi^c\dot\phi^2,
\end{eqnarray}
where we have defined $C_r=(\pi^2/30)g_*$. As in the constant-roll dynamics $\dot\phi=\dot\phi_0e^{-3\beta N}$ (assuming $\beta$ to be positive for the time being), we can write the above equation as 
\begin{eqnarray}
H=\left(\frac{C_\Upsilon M^{1-p-c}}{4C_r}T^{p-4}\dot\phi_0^2\right)\phi^ce^{-6\beta N}.
\end{eqnarray}
This situation is quite peculiar because now $H$ has become an explicit function of $\phi$. In a general situation, either slow-roll or constant-roll with constant $Q$ and where $\Upsilon$ is a function of $T$ alone, the $\phi$ dependence of $H$ comes from the potential of the inflaton field appearing in the first Friedmann equation (see Eq.~(\ref{Hsquare-w})). However, in the present case, the dependence of $H$ is not due to the inflaton potential but solely due to the constant-roll condition imposed on the constant radiation bath with nearly constant temperature. Thus, the evolution of $H$ according to the first Friedmann equation and according to the above equation will not be the same. Proceeding with the above form of $H$ and knowing that $H(d\phi/dN)=\dot\phi$, we can replace the form of $H$ from the above equation to obtain 
\begin{eqnarray}
\frac{d\phi}{dN}=\left(\frac{4C_rT^{4-p}}{C_\Upsilon M^{1-p-c}}\right)\frac{1}{\phi^c\dot\phi_0}e^{3\beta N}.
\end{eqnarray}
One can integrate this equation assuming $T$ to be a constant to obtain $\phi$ as a function of $N$ as 
\begin{eqnarray}
    \phi^{c+1}=\Bar{\phi}^{c+1}+\left(\frac{4(c+1)}{3\beta}\frac{C_r T^{4-p}}{C_\Upsilon M^{1-p-c}\dot\phi_0}\right)\left(e^{3\beta N}-1\right), \nonumber\\
\end{eqnarray}
where $\Bar{\phi}$ is the value of $\phi$ at $N=0$. Again, as this form of $\phi$ is not produced from the background Friedmann equations, it will not satisfy the WI inflaton equation of motion with constant-roll condition imposed on it (see Eq.~(\ref{KG-Warm-CR})). Thus the whole dynamics becomes inconsistent as soon as we impose the thermal stability condition, through Eq.~(\ref{rho-T}) and Eq.~(\ref{rho-approx}), on such constant-roll WI models where $\Upsilon$ is a function of $\phi$. This indicates that such constant-roll WI models cannot be treated near thermal equilibrium condition. The same argument holds for the negative $\beta$ cases as well. 

A quite common dissipative coefficient used in WI models is of the form $\Upsilon\propto T^3/\phi^2$ \cite{Berera:2008ar, Bastero-Gil:2010dgy, Bastero-Gil:2012akf}, where WI takes place in weak dissipative regime. According to our analysis, such a model cannot be treated with constant-roll dynamics while maintaining thermal equilibrium of the system.

\section{Discussion and Conclusion}
\label{conclusion}

In this paper we derived the conditions to obtain constant-roll WI models where (a) constant-roll dynamics drives at least 60 $e-$folds of inflation, (b) gracefully exits a constant-roll phase, and (c) the thermalization condition ($T>H$) is maintained throughout the constant-roll phase as it happens during standard slow-roll dynamics of WI. In doing so, we showed that both positive and negative values of $\beta$ are allowed in constant-roll WI. However, the thermalization condition forces one to restrict to values of $|\beta|$ much smaller than unity, which signifies that the dynamics of the constant-roll phase is not significantly different from a slow-roll evolution. In constant-roll CI, too, one requires to consider such small values of $\beta$ so that the predictions of the theory are in tune with the observations \cite{Motohashi:2017aob}, and thus the dynamics does not evolve into non-slow-roll regimes there as well. These conditions on $\beta$ in constant-roll CI is obtained from perturbation analysis. However, in constant-roll WI, the background evolution itself restricts the dynamics to evolve near slow-roll regime. We first consider constant-$Q$ toy models of WI to derive the dynamics of constant-roll WI. We numerically checked the models for both positive and negative values of $\beta$ in both weak and strong dissipative regimes. We then showed that the constant-$Q$ toy model analysis can be naturally extended to more realistic models of WI where the dissipative coefficient $\Upsilon$ is a function of $T$ alone. However, the most important result of our analysis is to show that WI models with dissipative coefficient $\Upsilon$ as a function of both $T$ and $\phi$ is untenable in thermal equilibrium once the constant-roll conditions are imposed. Therefore, such WI models cannot be realized with constant-roll conditions. More importantly, this analysis clearly indicates that the constant-roll dynamics can in certain circumstances be significantly different from slow-roll dynamics in WI. We further want to point out that in our previous analysis exploring a transient ultra-slow-roll phase in WI \cite{Biswas:2023jcd}, we noted that models with $\Upsilon$ as a function of $T$ alone cannot maintain thermal equilibrium even for a few $e-$foldings, whereas models with $\Upsilon$ as a function of $T$ and $\phi$ both can successfully have a transient ultra-slow-roll phase of a few $e-$foldings while maintaining near thermal equilibrium condition. However, what we have noticed in this paper is that, unlike in the ultra-slow-roll case, models with $\Upsilon(T)$ are consistent with constant-roll whereas models having $\Upsilon(\phi,T)$ are not suitable for constant-roll WI when thermal equilibrium is maintained. This also indicates that, unlike CI, the constant-roll dynamics is not a generalization of the ultra-slow-roll dynamics in WI. 

From the above discussion some key features of constant-roll WI emerges which we state below:
\begin{enumerate}
\item We showed that constant-roll WI occurring in thermal equilibrium is similar to a slow-roll WI phase if we purely want to describe an inflationary phase by the behaviour of the Hubble slow-roll parameters (both $\epsilon_1$ and $\epsilon_2$ being smaller than unity throughout the evolution), but this similarity is not deep rooted. Thermalization of constant-roll WI yields restrictions on the system (by constraining the value of $|\beta|$), whereas in standard slow-roll WI the thermalization of the radiation bath occurs naturally.

\item The constant-roll dynamics also differes from the slow-roll dynamics in the sense that the potential for the constant-roll is determined by the dynamics, unlike in the case of slow-roll where one is free to choose among any flat potentials which yield observarionally viable models. This is true for both CI and WI constant-roll models. 

\item Lastly, the fact that constant-roll WI must happen maintaining  thermal equilibrium yields further important restrictions on the theory by constraining the form of the dissipative coefficient. Thermal equilibrium rules out cases of constant-roll WI where the dissipation coefficient is a general function of $T$ and $\phi$. 
\end{enumerate}

It would be interesting to check whether such constant-roll WI models are universal attractors or not. Also, analysis of cosmological perturbations for such constant-roll WI models is not done in this paper. We defer both these issues for future projects in the hope that such analysis can help figure out whether the positive or negative $\beta$ is preferred.

\acknowledgements

The work of S.D. is supported by the Start-up Research Grant (SRG) awarded by Science and Engineering Research Board (SERB) [Grant No: SRG/2023/000101/PMS], Department of Science and Technology (DST), Government of India. S.D. is also thankful to Axis Bank and acknowledges the financial support obtained from them which partially supports this research. S.D. thanks Rudnei Ramos for many useful discussions on Warm Inflation on various occasions.



\end{document}